\documentclass[aps,twocolumn,showpacs]{revtex4-1}

\usepackage{graphicx}
\usepackage{dcolumn}
\usepackage{bm}

\usepackage[version=3]{mhchem}
\usepackage{epstopdf}

\begin{document}

\author{James L. Webb}
\email{j.l.webb@bath.ac.uk}
\author{Lewis S. Hart}
\author{Daniel Wolverson}
\affiliation{Centre for Nanoscience and Nanotechnology, Department of Physics, University of Bath, Bath BA2 7AY, United Kingdom}
\author{Chaoyu Chen}
\author{Jose Avila}
\author{Maria C. Asensio}
\email{maria-carmen.asensio@synchrotron-soleil.fr}
\affiliation{Synchrotron SOLEIL, Saint Aubin, and Universit\'{e} Paris-Saclay, BP 48 91192 Gif-sur-Yvette, France}

\date{\today}

\title[]{Electronic bandstructure of ReS$_2$ by high resolution angle resolved photoemission spectroscopy}


\begin{abstract}
The rhenium-based transition metal dichalcogenides (TMDs) are atypical of the TMD family due to their highly anisotropic crystalline structure and are recognized as promising materials for two dimensional heterostructure devices. The nature of the band gap (direct or indirect) for bulk, few and single layer forms of ReS$_2$ is of particular interest, due to its comparatively weak inter-planar interaction. However, the degree of inter-layer interaction and the question of whether a transition from indirect to direct gap is observed on reducing thickness (as in other TMDs) are controversial. We present a direct determination of the valence band structure of bulk ReS$_2$ using high resolution angle resolved photoemission spectroscopy (ARPES). We find a clear in-plane anisotropy due to the presence of chains of Re atoms, with a strongly directional effective mass which is larger in the direction orthogonal to the Re chains (2.2~$m_e$) than along them (1.6~$m_e$). An appreciable inter-plane interaction results in an experimentally-measured difference of $\approx$100-200 meV between the valence band maxima at the Z point (0,0,$\frac{1}{2}$) and the $\Gamma$ point (0,0,0) of the three-dimensional Brillouin zone. This leads to a direct gap at Z and a close-lying but larger gap at $\Gamma$, implying that bulk ReS$_2$ is marginally indirect. This may account for recent conflicting transport and photoluminescence measurements and the resulting uncertainty about the nature of the band gap in this material.
\end{abstract}

\maketitle

\section{Introduction}

The transition metal dichalcogenides (TMDs) are a class of material that can form thin sheets down to a single monolayer, analogous to to graphene but consisting of compound semiconducting materials rather than a single atomic species. It is these semiconducting properties that have attracted considerable recent interest with regard to fabricating and controlling new devices from stacked two-dimensional (2D) layered materials, the van der Waals heterostructures\cite{Geim2013}. The fundamental properties of other TMD materials such as MoS$_2$ and WS$_2$ have been intensively studied in recent years \cite{PhysRevLett.105.136805, ANGE:ANGE201000009, Coleman568} with a view towards creating novel electro-optic devices \cite{PhysRevB.90.125440}.

Of these materials ReS$_2$ is of particular interest due to several properties. Its optical \cite{doi:10.1063/1.354268,doi:10.1021/acsphotonics.5b00486} and electrical transport properties are highly anistropic, in particular with higher mobility in certain in-plane crystallograpic directions as determined by electrical measurements using bulk-like flakes, \cite{2015NatCo...6E6991L,doi:10.1021/acsnano.5b04851} making it an interesting material for the fabrication of FETs and polarization-sensitive photodetectors \cite{ADFM:ADFM201500969, 2015NatCo...6E6991L,Liu2016}. In this context, ReS$_2$ has the advantage of being stable under ambient conditions, unlike some similar 2D materials \cite{2053-1583-2-1-011002}. Next, spin-orbit coupling is important in ReS$_2$ though the presence of inversion symmetry means that spin-orbit splitting in unperturbed layers of any thickness is zero and may be manipulated via doping or gating. Finally, unlike many other TMD materials \cite{doi:10.1021/nl302584w,doi:10.1021/nl5045007} it has been proposed there is no transition from indirect to direct band gap with reduced thickness, meaning that the distinction between mono- and few-layer structures is not crucial for device concepts\cite{Tongay2014}. If true, this could be advantageous in terms of building efficient optical devices.

\begin{figure}
\includegraphics[width=9cm]{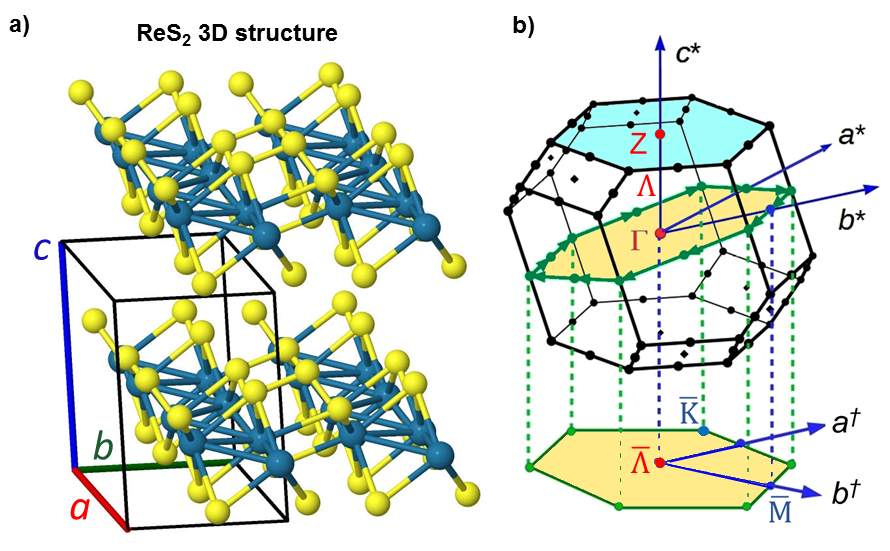}\
\caption{The atomic structure of ReS$_2$: (a) Re atoms in blue and S in yellow with the unit cell shown by the vectors \textbf{\textit{a}}, \textbf{\textit{b}}, \textbf{\textit{c}}. Chains of Re atoms run along the in-plane vector taken here as \textbf{\textit{a}}, with the second in-plane vector \textbf{\textit{b}} at about 120$^\circ$ to the chain direction. (b) The conventional triclinic Brillouin zone of ReS$_2$, showing the reciprocal lattice vectors \textbf{\textit{a}}$^{*}$, \textbf{\textit{b}}$^{*}$ and \textbf{\textit{c}}$^{*}$, with the latter passing through the labeled $\Gamma$ and Z points along the direction $\Lambda$. Two planes which pass through $\Gamma$ (the $\Gamma$ plane) and Z (the Z plane) and other high symmetry points are shaded; in our ARPES experiments, we effectively measure a projection of the 3D Brillouin zone onto the plane $k_z$=constant. This is shown schematically at the bottom of (b) by the irregular hexagon which is the projection of the $\Gamma$ plane. \textbf{\textit{a}}$^{\dagger}$ and \textbf{\textit{b}}$^{\dagger}$ represent the projections of the reciprocal lattice vectors in the $k_z$=constant plane, $\overline{K}$, $\overline{M}$ are the $k_z$=constant projections of their respective high symmetry points, and and $\overline{\Lambda}$  is the projection of the general point $\Lambda$.}
\label{figure_1}
\end{figure}

Figure~\ref{figure_1}(a) shows a schematic model of the atomic structure of ReS$_2$ with the lattice vector directions \textbf{\textit{a}},\textbf{\textit{b}} and \textbf{\textit{c}} indicated. Here \textbf{\textit{a}} lies in the direction of the Re chains and \textbf{\textit{b}} at about 120$^\circ$ to them. This structure has been confirmed experimentally by scanning probe microscopy studies \cite{doi:10.1021/ja00096a048}, indirectly by Raman spectroscopy \cite{Tongay2014,doi:10.1021/acs.nanolett.5b00910} and by single crystal x-ray diffraction by Lamfers \textit{et al.} \cite{LAMFERS199634} with a triclinic structure and in-plane lattice parameters $a$=6.352 and $b$=6.446~\AA. The question of whether the unit cell contains one or two layers stacked along the out-of-plane $c$ axis has been resolved in favor of a single layer, giving four formula units per unit cell\cite{Ho1999} with $c=6.403$~{\AA} to $6.461$~\AA\cite{Murray1994}.

In performing the ARPES measurements we effectively probe the projection of the $k$-points in the 3D Brillouin zone (BZ) of the material onto a quasi-2D flat plane ($k_z$=constant; the value of $k_z$ is determined by the choice of excitation photon energy). This is exemplified in Fig.~\ref{figure_1}(b) for points in the full 3D BZ in a plane passing through the $\Gamma$ point (shaded yellow), which are projected onto the measurement plane to produce a quasi-2D Brillouin zone in $k_x$, $k_y$. This quasi-BZ is shown at the bottom of Fig.~\ref{figure_1}(b) and is an irregular hexagon centered on $\overline{\Lambda}$. Here we show the projection of the lattice vectors \textbf{\textit{a}$^{*}$} and \textbf{\textit{b}$^{*}$} onto the measurement plane as \textbf{\textit{a}$^{\dagger}$} and \textbf{\textit{b}$^{\dagger}$}. If the bulk material were thinned to a 2D monolayer, this quasi-BZ would ultimately represent a good approximation to the 2D BZ as used in other work, for example Tongay et al. \cite{Tongay2014} to calculate the properties of monolayer ReS$_2$. We label directions $\overline{M}$ and $\overline{K}$ following this and other work and in order to show the direction of measurement of the ARPES data. It is necessary to distinguish between directions $\overline{M}_1$..$\overline{M}_3$ and $\overline{K}_1$..$\overline{K}_3$ since, in this triclinic material, none of these are related by symmetry. The directions \textbf{\textit{b}$^{*}$} and \textbf{\textit{b}$^{\dagger}$} are orthogonal to the real space vector \textbf{\textit{a}}, so that directions $\overline{M}_1$ and $\overline{K}_2$ are exactly orthogonal and approximately parallel, respectively, to the rhenium chains of ReS$_2$.

Although electrical transport and optical absorption measurements have given some indirect insight into the ReS$_2$ bandstructure, along with \textit{ab initio} calculations of the material properties \cite{PhysRevB.60.15766,Tongay2014,Dileep2016}, it is evident that gap size as well as the  direct or indirect character of the gap is particularly sensitive to the details of any computational model (for instance, the choice of pseudopotential, and whether spin-orbit coupling is included). Consequently, the direct, accurate determination of the electronic bandstructure is necessary and timely. This is a task for which angle resolved photoemission spectroscopy (ARPES) is ideally suited. However, two primary difficulties exist in performing these measurements: they must be performed on a clean surface, ideally a crystal cleaved under ultra-high vacuum conditions, and the crystal facets of the material must be larger than the spot size of the X-ray beam in order to obtain clear, monocrystalline data. Recent advances in TMD handling and advances towards micro and nano-ARPES systems with beam sizes on the 100~nm scale mean these problems can now be overcome.

In this work we present the main results of our direct determination of the valence bandstructure of bulk ReS$_2$ using nano-ARPES, mapping the photoemission intensity in directions along and perpendicular to the Re atomic chains of the material and measuring also constant binding energy contours throughout large portions of the full 3D Brillouin zone. Our findings show the effect of the in-plane anisotropy on the electronic dispersions and the van der Waals interactions between planes, revealing the electronic dispersion perpendicular to the constitutive layers.
Most importantly, we also have found significant differences between the electronic structure at the valence band maxima at the $Z$ (0,0,$\frac{1}{2}$) and the $\Gamma$ (0,0,0) points. We find there is a high-lying valence band maximum at the $Z$ point moving to a lower-lying one at $\Gamma$, in agreement with our calculated band structure. In conjunction with our calculated conduction band minima, this implies a direct band gap at the $Z$ point with an indirect gap at or near the $\Gamma$ point. All these observations together provide an explanation for the anisotropy observed in electrical and optical measurements and the uncertainty regarding the direct or indirect nature of the ReS$_2$ band gap.

\begin{figure*}
\includegraphics[width=14cm]{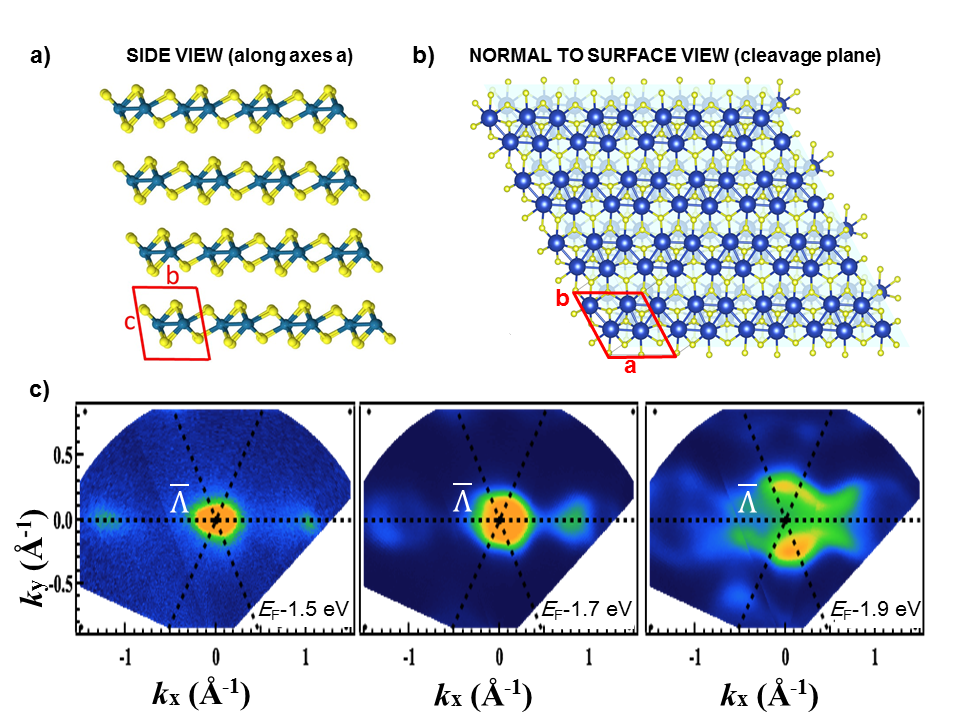}\
\caption{Side (a) and normal (b) view of the ReS$_2$  structure; (c) constant energy surfaces going down in the valence band from 1.5~eV to 1.9~eV below the Fermi energy $E_F$ and centered at the $\overline{\Lambda}$ point as discussed in the text. Significant asymmetry is observed between the direction \textbf{\textit{a}} along the chains (which are parallel to $k_y$) and that perpendicular to them ($k_x$, or \textbf{\textit{b}$^{\dagger}$}). The maximum of the valence band (VBM) for this value of $k_z$ (see text) appears at $\overline{\Lambda}$.}
\label{figure_2}
\end{figure*}

\section{Methods}

Samples were commercially grown via Bridgman single crystal growth by 2D Semiconductors USA and were confirmed 99.9995$\%$ pure using secondary ion mass spectrometry. We performed prior studies on the crystals using Raman spectroscopy in order to confirm their phase and high crystal quality\cite{Hart2016}.

Nano-ARPES measurements were performed using the k-microscope at the \textit{ANTARES} beamline at the Soleil synchrotron, Paris, with a spot size of 100~nm. At a photon energy of 100~eV, this beamline has an angular resolution of $\sim$0.2$^\circ$ and an energy resolution of $\sim$10~meV. The advantage of the nano-ARPES technique for the study of ReS$_2$  is that the X-ray beam spot size is smaller than the size of the crystallites (generally a few microns to tens of microns, as determined by optical microscopy), ensuring that the measured dispersion is obtained only from a single facet, which is important due to the high degree of direction-dependent anisotropy in the band structure. The samples were prepared by cleaving \textit{in-situ} under UHV conditions (pressure $<1\times 10^{-10}$~mbar). A nano-positioning system was used to locate clean and flat areas of the sample with maximum ARPES intensity with measurements performed at photon energies from 95-180~eV. The sample was rear-cooled using liquid nitrogen to approximately 100~K in order to reduce thermal noise.

DFT calculations were performed using the Quantum Espresso package \cite{0953-8984-21-39-395502} to perform structural relaxation and obtain total energy and bandstructure simulations. We focus here on results obtained using a non-relativistic Perdew Burke Ernzerhof (PBE) generalised gradient approximation (GGA) exchange-correlation functional\cite{PhysRevLett.77.3865} but we also explored the use of a fully-relativistic Perdew Zunger (PZ) local density approximation (LDA) functional\cite{PhysRevB.23.5048} with projector augmented wave (PAW) pseudopotentials generated by QE and PSLibrary46. The GGA and LDA results are compared to the experimental data in the Supplemental material, Figures S1 and S2 respectively \cite{SM}. The valence of Re was taken as 15 (configuration $5s^2 5p^6 5d^5 6s^2$). Kinetic energy cutoffs were 70~Ry (816~eV) and Monkhorst-Pack $k$-point meshes of $12\times12\times12$ were used with a single 12-atom unit cell.

\section{Results}

First, we consider how the observed constant energy maps in the $k_z=$constant plane reflect the crystal symmetry; Fig.~\ref{figure_2} shows views of the crystal structure looking along the \textbf{\textit{a}} and parallel to the \textbf{\textit{c}} directions respectively. In Fig.~\ref{figure_2}(c), we show a constant energy surface plot of the ARPES signal intensity probing a set of binding energies E$_b$ moving downwards in energy from the Fermi energy ($E_F$) and recorded by illuminating the samples with photons of energy h${\nu}$ = 100~eV. This excitation energy was chosen since it gives the optimum transmission of the zone plate used to focus the excitation beam. As we shall see below, this excitation energy means that the plane probed intersects the \textbf{\textit{c}$^{*}$} axis at a point lying on the line $\Gamma$-Z. Following the notation of Fig.~\ref{figure_1}(b), we label a general point of this type as $\overline{\Lambda}$; this point is the origin ($k_x = k_y = 0$) of the 2D quasi-Brillouin zone, and thus is the origin of the binding energy contour plots for a given excitation energy.

\begin{figure*}
\includegraphics[width=14cm]{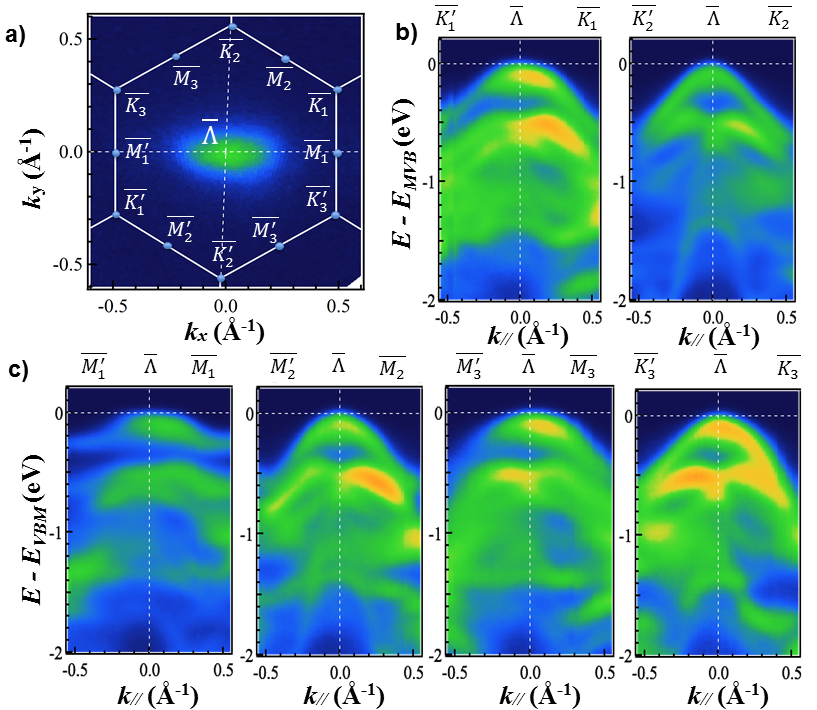}\
\caption{(a) shows the constant energy surface in the valence band at 1.5~eV below the Fermi energy $E_E$ as shown in Fig.~\ref{figure_2}(c) with the labels of the high symmetry points projected onto the ($k_x$,$k_y$) plane.  Panels (b) and (c) show high resolution ARPES plots recorded using an excitation photon energy of h$\nu=100$ eV along the six inequivalent $\overline{K'}$-$\overline{\Lambda}$-$\overline{K}$ and  $\overline{M'}$-$\overline{\Lambda}$-$\overline{M}$ directions.}
\label{figure_3}
\end{figure*}

Using a gold sample \emph{in situ} in the ARPES system, the Fermi edge of its density of states was precisely determined; this experimental Fermi energy is shared by the ReS$_2$ sample as they have a common ground potential. We find that the Fermi level of the semiconductor sample is 1.5 eV above the valence band maximum, putting a lower bound on the single-particle band gap of 1.5~eV (unoccupied bands are not recorded by ARPES, so the conduction band minimum was not recorded here). This value is close to the lowest-lying excitonic band gap of $E_{1}^{ex}=1.55$~eV recorded at the same temperature (100~K) \cite{Ho1998}, showing the $n$-type character of our material. The doping state of ReS$_2$ is dependent on the details of the crystal growth with $p$-type material also being possible depending on the vapor transport method used\cite{LEICHT1987531,doi:10.1063/1.354268}. Consequently, the ARPES constant energy plots of Fig.~\ref{figure_2}(c) can be labeled at this local maximum of the valence band with binding energies of 1.5 eV, 1.7 eV and 1.9 eV below the Fermi level.

The plots of Fig.~\ref{figure_2}(c) clearly indicate a ``wavy'' shape of the contours related to the in-plane chains of Re atoms sketched in Fig.~\ref{figure_2}(b). This implies a marked difference between the dispersion in the direction along the Re chains and that perpendicular to them, with a more abrupt drop in the valence band energy along the chains. Such anisotropy directly affects the effective mass and hence the mobility in the two perpendicular directions. This aspect will be studied in detail below, where we obtain representative effective masses in both directions from our ARPES data. However, it is important to note that this in-plane anisotropy is not exclusive to these two directions. Fig.~\ref{figure_3} indicates that the band structure along different $\overline{M'}$-$\overline{\Lambda}$-$\overline{M}$ directions is dissimilar, confirming that the in-plane anisotropy is not restricted to the directions along and perpendicular to the chains of Re atoms. The same remark can be made about  the  electronic dispersions along the $\overline{K'}$-$\overline{\Lambda}$-$\overline{K}$ directions, which are also not equivalent, even though the tops of the bands in all directions are centered at the surface $\overline{\Lambda}$ projection, as shown in Fig.~\ref{figure_3}(a). We distinguish in Fig.~\ref{figure_3} between points either side of $\overline{\Lambda}$ (e.g, $\overline{M'}$ and $\overline{M}$) because there is inversion symmetry only through the \emph{true} $\Gamma$ point and so these pairs are not related by real reciprocal lattice vectors. Indeed, it can be seen in some of the ARPES plots of Fig.~\ref{figure_3}(b) and (c) that the dispersion is not exactly symmetrical about $\overline{\Lambda}$ for this reason.

\begin{figure}
\includegraphics[width=7.5cm]{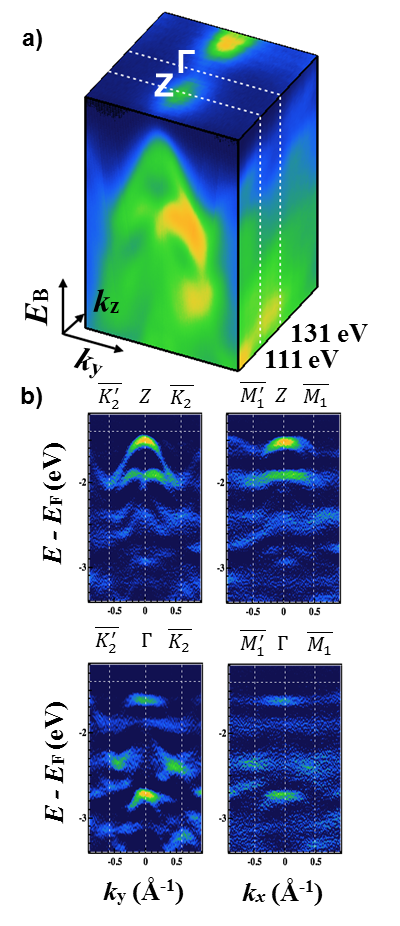}\
\caption{(a) 3D ARPES intensity plots along the  $k_z$, $k_y$ versus binding energy E$_B$. The $k_z$ periodicity shows that at photon energies of 131 eV and 111 eV, the ARPES mapping probe planes containing the $\Gamma$ and the $Z$ symmetry points, respectively. From this experimental determination we have estimated the inner potential V$_in$ = $16\pm$2eV. (b) Second derivatives of ARPES intensity taken at $Z$ and $\Gamma$ along the $\overline{K}_2$ and $\overline{M}_1$ directions.  In both cases, we see a sharper peak at $Z$, compared to the flatter bands in $\Gamma$, particularly in the $\overline{M}_1$ direction perpendicular to the Re chains. We see a difference in energy between the top of the valence band at $Z$ and that at $\Gamma$ of approximately ~150-200 meV. }
\label{figure_4}
\end{figure}

Thus far, we have measured the bands only at an arbitrary point along \textbf{\textit{c}$^{*}$} (corresponding to the excitation energy of 100~eV) and this limits our analysis. Consequently, it is important to probe the limits of the 3D Brillouin zone, at the $\Gamma$ and $Z$ points. This can be achieved in ARPES by varying the excitation photon energy; the systematic study of the energy dependence of the ARPES signals allows the mapping of the whole ReS$_2$ 3D BZ and a precise analysis then provides the exact location of the planes intersecting the ${\Gamma}$ and $Z$ points of the ReS$_2$ 3D Brillouin zone, defined in Fig.~\ref{figure_1}(b). In brief, in order to obtain experimentally the perpendicular dispersion of the bands, ARPES plots are recorded systematically by varying $k_z$, scanning through the $\Gamma$ point at  $k_z$=0 and the $Z$ point at $k_z =|c^{*}|/2$. Figure~\ref{figure_4}(a) depicts the out-of-plane dispersion of the bands (that is, along $k_z$), obtained as the incident photon energy is varied. The results plotted in Fig.~\ref{figure_4}(a) demonstrate that photon energies of 131 eV and 111 eV correspond to the $\Gamma$  and $Z$ points respectively, consistent with initial electron state momenta that are integer and half-integer multiples of $|c^{*}| = 1.03$~\AA$^{-1}$, if we assume a value of the inner potential for ReS$_2$ of $V_{in}=16\pm2$eV. This potential is conventionally used to represent the effects of the non-conservation of photoelectron momentum normal to the emitting surface\cite{Huefner2003} and the value we obtain is consistent with those of the similar TMDs WSe$_2$ and ReSe$_2$\cite{Finteis1997,Hart2017}.

Figure~\ref{figure_4}(b) shows plots of the the second derivative of the dispersion for the two most representative directions, along $k_x$ and $k_y$ ($\overline{M}_1$ and $\overline{K}_2$) with excitation energies selected from the dataset of Figure~\ref{figure_4}(a) so that the selected planes pass through the $Z$ and $\Gamma$ symmetry points (upper and lower panels respectively). The in-plane anisotropy of the dispersions along the $k_x$ and $k_y$ directions is very noticeable. Interestingly, a distinct inequivalence between the electronic band dispersion at $Z$ and $\Gamma$ points is also observed. A typical, very flat top to the valence band appears for $\overline{M'}_1$-$\Gamma$-$\overline{M}_1$, while more dispersive bands characterize the VBM at the $Z$ point.
We find that the binding energy of the  VBM at the $Z$ point is lower than at $\Gamma$ with a  difference of 100-200~meV. The 3D plot of the same data in $k_x$ and $k_y$ versus E$_B$ shows a clear, single peak in the valence band at $Z$, as might be characteristic of a direct bandgap transition at this point. In previous recent indirect measurements and \textit{ab inito} calculations, a direct bandgap at $\Gamma$ had been proposed\cite{Tongay2014,PhysRevB.92.115438} though no calculations of the valence band throughout the whole Brillouin zone have been reported.

\begin{figure*}
\includegraphics[width=15cm]{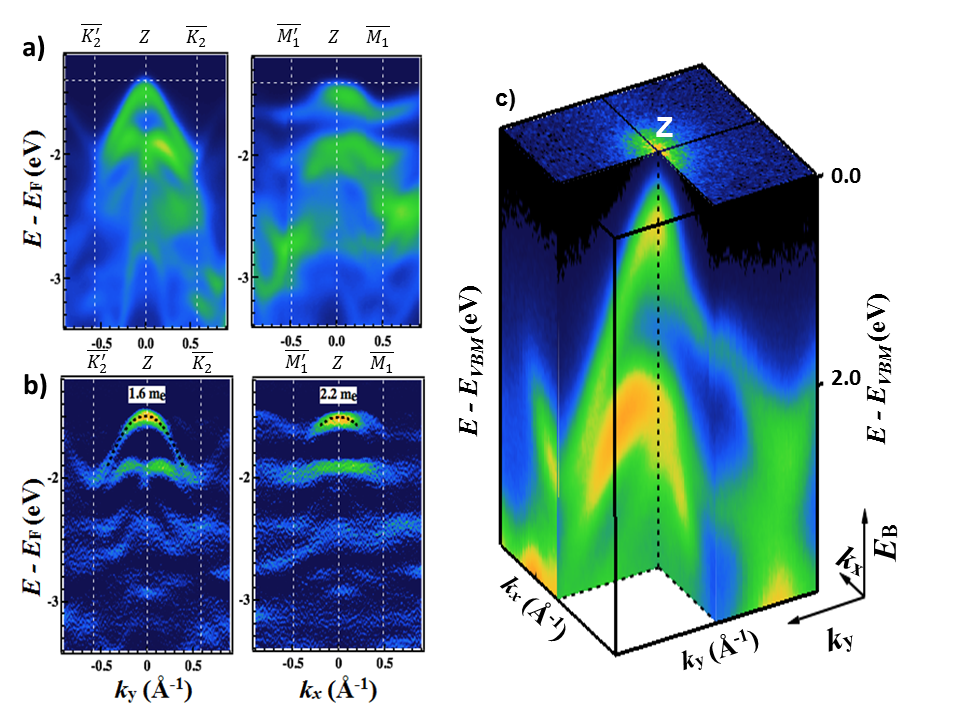}\
\caption{a) ARPES plots along the $\overline{K'}_2$-$Z$-$\overline{K}_2$ and $\overline{M'}_1$-$Z$-$\overline{M}_1$ directions are plotted together in (b) with the second derivative ARPES bands and the fitting curves determining the effective mass in both directions. (c) the 3D valence band dispersion along the $k_x$ and $k_y$ directions.}
\label{figure_5}
\end{figure*}

To determine quantitatively some key details of the electronic structure of ReS$_2$, Figure~\ref{figure_5} shows the ARPES intensity recorded along $\overline{K'}_2$-$Z$-$\overline{K}_2$ and $\overline{M'}_1$-$Z$-$\overline{M}_1$ directions. Despite the low symmetry in the plane, the band dispersions are approximately symmetric about the $Z$ point, as Fig.~\ref{figure_5}(a)-(c) show. The fitted parabolae (dotted black lines in Fig.~\ref{figure_5}(b) allow a precise estimation of the degree of in-plane anisotropy in the valence band. Here, effective valence band masses of 1.6$\pm$0.3~$m_e$ and 2.2$\pm$0.7~$m_e$ (where $m_e$ is the free electron mass) have been directly determined along and perpendicular to the Re atomic chains respectively (in the Supplemental material, Figure S1, we show fits bracketing the values above superimposed on the experimental data \cite{SM}). These values lie in the typical range for TMD materials \cite{Rasmussen2015}.

Recently, the electrical transport properties of $n$-type field effect transistor devices have been reported to be strongly anisotropic \cite{2015NatCo...6E6991L, Corbet2015, Zhou2017}, and the conduction band electron mobility was found to be about three times larger along the rhenium chains compared to the direction perpendicular to them \cite{2015NatCo...6E6991L}. In the Supplemental material, Figure S4, we give the conduction band effective masses derived from DFT calculations, which reflect this anisotropy \cite{SM}. The valence band masses given above would also lead to a similar degree of anisotropy in the hole mobility even if all other parameters determining the phonon-limited mobility (deformation potential constant, elastic modulus) were isotropic. However, no electrical measurements of hole mobility in ReS$_2$ are yet available for comparison.

\begin{figure}
\includegraphics[width=8cm]{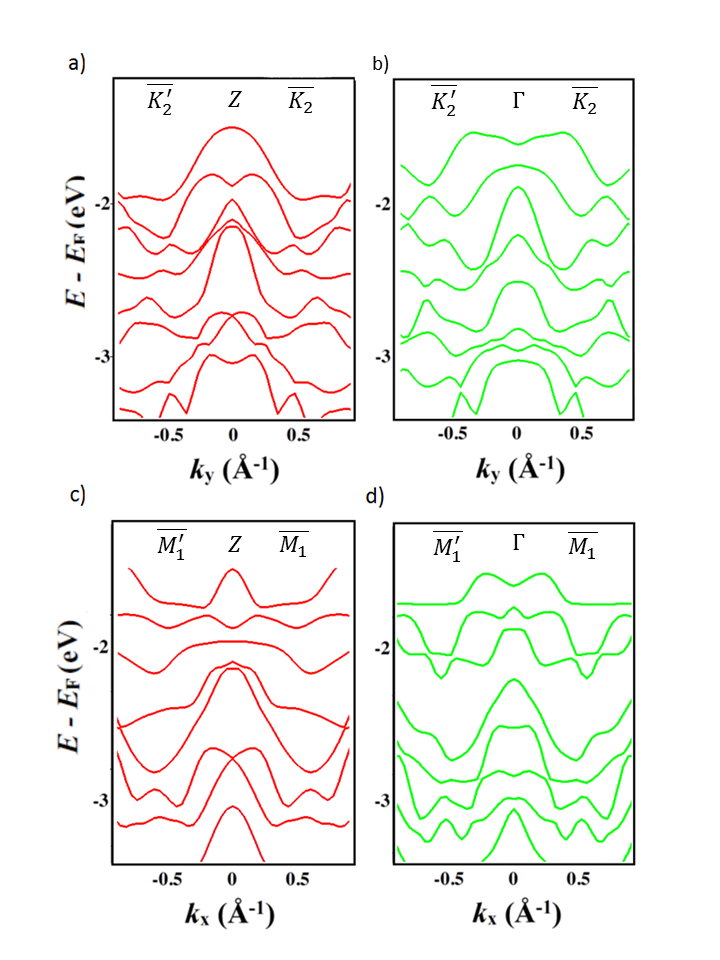}\
\caption{Calculations of the valence bands measured by ARPES in Figure~\ref{figure_4} along the directions: a) $\overline{K'}_2$-$Z$-$\overline{K}_2$; b) $\overline{K'}_2$-$\Gamma$-$\overline{K}_2$; c) $\overline{M'}_1$-$Z$-$\overline{M}_1$; d) $\overline{M'}_1$-$\Gamma$-$\overline{M}_1$ by DFT using a PBE functional (parameters given in method), showing the same anisotropy with respect to the Re chain direction.}
\label{figure_6}
\end{figure}

To aid in interpretation of the data, we performed simulations of the present data using DFT calculations using parameters detailed in the Method section; some results are shown in Figure~\ref{figure_6}. We obtain qualitative agreement with the experimental data in both the $\overline{K}_2$ and $\overline{M}_1$ directions, with the calculations matching the uppermost experimental bands best for  $\overline{K}_2$-$Z$ and worst for $\overline{M}_1$-$\Gamma$. To facilitate comparison of these simulations to experiment, figures S1 and S2 of the Supplemental material show the simulations superimposed on the ARPES data for calculations using both the PBE functional (as shown in Figure 6; Fig. S1) and a fully relativistic LDA functional (Fig. S2) \cite{SM}.

Importantly, in both cases we replicate the anisotropy observed in experiment with respect to the Re chain direction. As an example, we focus on the dispersion along $\overline{K'}_2$-$\Gamma$-$\overline{K}_2$, Fig.~\ref{figure_6}(a), where the highest energy valence band has a strongly peaked and approximately parabolic form whilst the next valence band down in energy has a distinctive double-peak structure; this is exactly as found in experiment, as shown in Figs.~\ref{figure_4}(b) and \ref{figure_5}(b). In these simulations, we did not take into account the slight curvature of the plane in $k$-space probed in ARPES that arises from the non-conservation of $k_z$\cite{Hart2017}; for these rather high excitation energies, we expect that this does not introduce a significant error but it may affect the comparison to experiment particularly for bands deeper in the valence band.

\begin{figure}
\includegraphics[width=8cm]{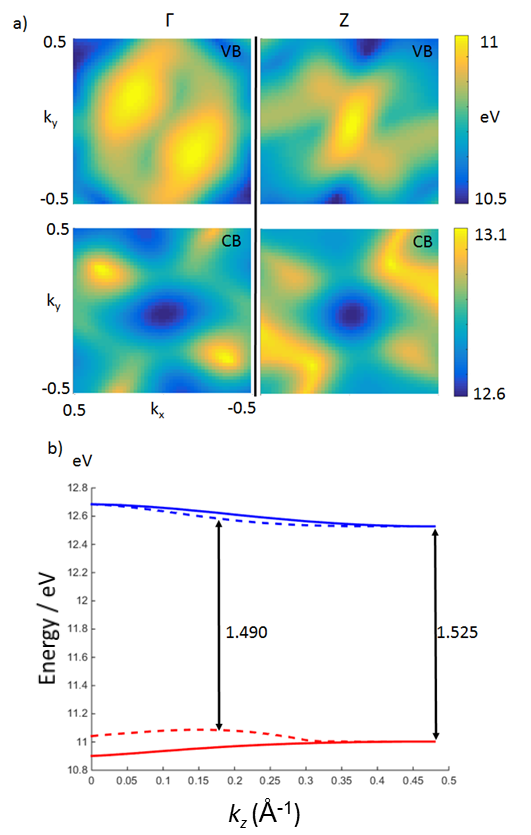}\
\caption{
Calculation of the valence band maximum (VB) and conduction band minimum (CB) at $Z$ and $\Gamma$ using a PBE functional. The calculations are performed in the shaded planes of Figure~\ref{figure_1}(b) and here $k_x$ and $k_y$ are in units of the reciprocal lattice vector \textbf{a$^{*}$}. We see a direct gap at $Z$ with an increasingly indirect gap towards $\Gamma$. (b) Calculation of the VB maximum and CB minimum as a function of $k_z$ for all $k_x$,$k_y$ (solid line) and restricted to the $\Gamma$-$Z$ direction $k_x$=0,$k_y$=0 (dashed line). We find the minimum indirect bandgap occurs at around $k_z$=0.18, lower in energy than the direct gap at $Z$. We calculate again a similar change in the VB maximum from $\Gamma$ to $Z$ as in experiment. }
\label{figure_7}
\end{figure}

Based on these simulations of the experimental data we used DFT to calculate the full band structure of the material including the lowest-lying conduction band states. Here, both PBE (GGA) and fully relativistic PZ (LDA) functionals were used in order to estimate the effects of spin-orbit coupling due to the high atomic number of rhenium. The results are qualitatively similar apart from the well-known underestimation of the band gap in the LDA, and so we focus here on the PBE (GGA) functional as used to produce the results shown in Fig.~\ref{figure_6}. Figure~\ref{figure_7}(a) shows the calculated energy of the highest energy VB and lowest energy CB states, with Fig.~\ref{figure_7}(b) showing the absolute CB minimum (CBM) and VB maximum (VBM) energies (dashed) and the local CBM and VBM values moving along the direction $\Gamma$ to $Z$, that is, as a function of $k_z$. We performed a full 3D calculation of the entire Brillouin zone in $k_x$, $k_y$ and $k_z$ to obtain this data. As noted above, conduction band effective masses obtained from parabolic fits to the DFT data at the Z point are given for the $k_x$ and $k_y$ directions in the Supplemental material (Fig. S4) \cite{SM}. From our calculations, we also obtained the electronic density of states for bulk ReS$_2$ (Supplemental material, Figure S3) which reproduces earlier calculations well \cite{Fang1997,Ho1999PRB}.

At $Z$ we obtain an estimate of the zero-temperature direct band gap as ~1.525 eV. We note that a broad range of experimental band gaps exist from previous work from around 1.6eV \cite{Tongay2014} down to around 1.3eV\cite{Ho1998,doi:10.1021/ja00096a048} (or lower in defect-rich materials or if defect states are probed\cite{PhysRevB.89.155433}). In our calculations, the band gap value obtained by the GGA is the most reliable and is in good agreement with experimental values\cite{Ho1998}. Interestingly, towards $\Gamma$ we find a transition to an indirect band gap. The conduction band minimum remains at ($k_x$,$k_y$)=0 (bottom left panel of Fig.~\ref{figure_7}a). We find the gap narrows at around $k_z$=0.18 to lower energy (1.49~eV) than at $Z$, as indicated by the arrow on Fig.~\ref{figure_7}(b). Moving from $Z$ to $\Gamma$ and calculating at ($k_x$,$k_y$)=0 (solid line on Fig.~\ref{figure_7}) we observe the valence band maximum to decrease by 125~meV for PBE (and 210meV for PZ). This is very similar to the 100-200~meV drop in the valence band maximum from $Z$ to $\Gamma$ observed experimentally and can be seen in Fig.~\ref{figure_4}.

This behavior may account for some of the uncertainty in the literature regarding the direct or indirect gap nature of bulk ReS$_2$, with other groups reporting electrical transport properties characteristic of indirect behavior in bulk samples \cite{Morpurgo2016,Ho1998} or PL peaks within this range close to the direct transition\cite{doi:10.1021/acsphotonics.5b00486}, whilst one group found a direct gap via electron energy loss spectroscopy of 1.42~eV at room temperature \cite{Dileep2016} , consistent with the direct gap at 100~K of $\sim$1.5~eV that we find \cite{Huang2013}. Unlike here, previous calculations have often not considered the full bulk BZ and have instead focused on 2D or quasi-2D simulations of ReS$_2$ where calculations have predicted the gap to be direct at the 2D $\Gamma$ point \cite{Tongay2014} in the monolayer form. As yet, only one brief study of ARPES measurements has been reported on a 2D monolayer of ReS$_2$ \cite{Liu2016}, and an important next step in this field will be to investigate few-layer structures in more detail. Use of a hybrid functional may also enhance the quality of the calculations. To confirm fully the nature of the band gap in the material it is necessary to measure the lowest states of the conduction band (potentially through alkali metal doping of the material). During the preparation of this paper, two more reports of ARPES studies of ReS$_2$ in bulk and thin-layer forms appeared on arxiv (both now published) which confirm the three-dimensional dispersion of the valence band structure \cite{Gehlmann2017} and, for the bulk material, indicate via Rb-doping that the band extrema are indeed located at $Z$ as proposed above \cite{Biswas2017}.

\section{Conclusions}
In conclusion, we have performed ARPES measurements on the transition metal dichalcogenide ReS$_2$ in its bulk form. The anisotropic curvature of the valence band in directions perpendicular and parallel to the Re atomic chains can account for the reported anisotropic electrical properties (effective mass and mobility) of the material. We find the valence band maximum at $Z$ to be higher in energy than that at $\Gamma$ and we obtain good agreement with DFT calculations of the band structure. Our calculations for the electronic bands over the whole bulk Brillouin zone predict a direct bandgap at $Z$ and a wider gap towards $\Gamma$ with a small shift of 100-200~meV in the valence band maximum between these two points. This may account for uncertainty in the literature as to the direct or indirect nature of the band gap in bulk ReS$_2$. Future work to measure the conduction band directly (for example by doping to move the Fermi level into the conduction band) is required to investigate this more fully and also to map out the band structure of two-dimensional monolayer ReS$_2$.

\begin{acknowledgments}
This work was supported by the Centre for Graphene Science of the Universities of Bath and Exeter and by the EPSRC (UK) under grants EP/G036101, EP/M022188 and EP/P004830; LSH is supported by the Bath/Bristol Centre for Doctoral Training in Condensed Matter Physics, grant EP/L015544. We thank the SOLEIL synchrotron for the provision of beam time; work at SOLEIL was supported by EPSRC grant EP/P004830/1. Computational work was performed on the University of Bath's High Performance Computing Facility. We thank Dr Philip King of St Andrews University for useful discussions and for communicating his data \cite{Biswas2017} prior to publication. M.C.A., J.A. and C.C. thank Young-Hee Lee and Matthias Bantzill for enlightening exchanges. The Synchrotron SOLEIL is supported by the Centre National de la Recherche Scientifique (CNRS). Data created during this research are freely available from the University of Bath data archive at DOI:10.15125/BATH-00331.
\end{acknowledgments}

\end{document}